\documentclass[12pt]{article}
\usepackage{amsmath}
\input{epsf}
\def\ltap{\raisebox{-.4ex}{\rlap{$\,\sim\,$}} \raisebox{.4ex}{$\,<\,$}}

\newcommand\as{\alpha_{\mathrm{S}}}

\def\beq{\begin{equation}}
\def\eeq{\end{equation}}
\def\beeq{\begin{eqnarray}}
\def\eeeq{\end{eqnarray}}

\def\to{\rightarrow}
\def\nn{\nonumber}

\begin{document}
\begin{titlepage}
\renewcommand{\thefootnote}{\fnsymbol{footnote}}
\begin{flushright}
    CERN--TH/2002-123   \\ hep-ph/0206052
     \end{flushright}
\par \vspace{10mm}

\begin{center}
{\Large \bf
Direct Higgs production and jet veto at hadron colliders
\footnote{Talk given by M. Grazzini at the XXXVIIth Rencontres de Moriond, QCD and Hadronic interactions, Les Arc1800, France}
\\[1.ex]
}

\end{center}
\par \vspace{2mm}
\begin{center}
{\bf Stefano Catani${}^{(a)}$~\footnote{On leave of absence 
from INFN, Sezione di Firenze, Florence, Italy.}, Daniel de Florian${}^{(b)}$
\footnote{Partially supported by Fundaci\'on Antorchas and Conicet.}
}
\hskip .2cm
and
\hskip .2cm
{\bf Massimiliano Grazzini${}^{(c,d)}$}\\

\vspace{5mm}

${}^{(a)}$Theory Division, CERN, CH-1211 Geneva 23, Switzerland \\

${}^{(b)}$Dep. de F\'\i sica, FCEYN, Universidad de Buenos Aires,
(1428) Pabell\'on 1 Ciudad Universitaria, Capital Federal, Argentina\\

${}^{(c)}$ Dipartimento di Fisica, Universit\`a di Firenze, I-50019 Sesto Fiorentino, Florence, Italy\\

${}^{(d)}$INFN, Sezione di Firenze, I-50019 Sesto Fiorentino, Florence, Italy

\vspace{5mm}

\end{center}

\par \vspace{2mm}
\begin{center} {\large \bf Abstract} \end{center}
\begin{quote}
\pretolerance 10000

We consider Higgs boson production through gluon--gluon fusion in hadron collisions, when a veto is applied on the transverse momenta of the accompanying hard jets.
We compute the QCD corrections to this process at NLO and NNLO, and present
numerical results at the Tevatron and the LHC.

\end{quote}

\vspace*{\fill}
\begin{flushleft}
     CERN--TH/2002-123 \\ June 2002 

\end{flushleft}
\end{titlepage}

\setcounter{footnote}{1}
\renewcommand{\thefootnote}{\fnsymbol{footnote}}


The search for the Higgs boson is one the main experimental challenges in high-energy physics.
In the near future this search will be carried out at hadron colliders,
the Tevatron 
and the LHC. 
LEP has established at 95$\%$ confidence-level that the mass $M_H$ of the
SM Higgs boson is larger than $114.1$~GeV \cite{Schwickerath:2002nf}, whereas
precision electroweak measurements
indicate that the Higgs boson should be light ($M_H\ltap 200$~GeV). 
At the Tevatron and the LHC,
various channels~\cite{Carena:2000yx,atlascms} can be exploited to search for the Higgs boson in this mass window.

Direct Higgs production
followed by
the decay $H\to W^*W^*,Z^*Z^*$ is relevant for a Higgs boson
with mass $140\ltap M_H \ltap 190$~GeV.
In particular, 
the decay mode $W^*W^*\to l^+l^-\nu {\bar \nu}$ is quite important
\cite{Carena:2000yx,atlascms,Dittmar:1997ss,Han:1999ma},
since it is cleaner than $W^*W^*\to l\nu jj$, and the decay rate $H\to W^*W^*$
is higher than $H\to Z^*Z^*$ by about one order of magnitude.

An important background for the direct Higgs signal 
$H\to W^*W^*\to l^+l^-\nu {\bar \nu}$
is $t {\bar t}$ production ($tW$ production is also important at the LHC), where
$t \to l{\bar \nu} b$, thus leading to $b$ jets with high $p_T$ in the final 
state. If the $b$ quarks are not identified, 
a veto cut on the transverse momenta of the jets accompanying
the final-state leptons
turns out to be essential, both at
the Tevatron \cite{Carena:2000yx,Han:1999ma} and at the LHC \cite{atlascms,Dittmar:1997ss},
to cut the hard $b$ jets arising from this background process.

Here we study the effect of a jet veto on the signal, or, more precisely,
on the cross section for
direct Higgs
production.
The events that pass the veto selection are those with
$p_T^{\rm jet} < p_T^{\rm veto}$, where $p_T^{\rm jet}$ is the transverse
momentum of any final-state jets, defined by a cone algorithm.
The cone size $R$ of the jets will be fixed
at the
value $R=0.4$.
More details of our study are presented elsewhere~\cite{Catani:2001cr}.

The vetoed cross section $\sigma^{\rm veto}$ at the c.m. energy $\sqrt s$
can be computed through the following factorization formula:
\vspace*{-3mm}
\begin{align}
\label{had}
\sigma^{\rm veto}(s,M_H^2;p_T^{\rm veto},R) &= 
\sum_{a,b} \int_0^1 dx_1 \;dx_2 \; f_{a/h_1}(x_1,\mu_F^2) 
\;f_{b/h_2}(x_2,\mu_F^2) \int_0^1 dz \;\delta\!\left(z -
\frac{\tau_H}{x_1x_2}\right) \nn \\
& \cdot \sigma_0\,z\;G_{ab}^{\rm veto}(z;\as(\mu_R^2), M_H^2/\mu_R^2;M_H^2/\mu_F^2;p_T^{\rm veto},R) \;,
\end{align}
where $\tau_H=M_H^2/s$, $f_{a/h}$ is the parton density of the colliding hadron
$h$, and $\mu_F$ and $\mu_R$ are the factorization and 
renormalization scales, respectively. 
The vetoed cross section  can also be written as 
\vspace*{-3mm}
\begin{equation}
\label{sigmaveto}
\sigma^{\rm veto}(s,M_H^2;p_T^{\rm veto},R) = \sigma(s,M_H^2) 
- \Delta \sigma(s,M_H^2;p_T^{\rm veto},R) \;\;,
\end{equation}
where $\sigma(s,M_H^2)$ is the inclusive cross section, and $\Delta \sigma$ is the `loss' in cross section
due to the jet-veto procedure.

The partonic cross section 
$\sigma_0 G_{ab}^{\rm veto}$ in Eq.~(\ref{had}) is
computable as a perturbative expansion in the QCD coupling $\as$.
Our calculation is performed by using 
the large-$M_{\rm top}$ approximation, but the Born cross section $\sigma_0$ 
is evaluated exactly.
At NLO the coefficient function $G^{\rm veto}_{ab}$ 
can be computed analytically \cite{Catani:2001cr}.
At NNLO we subtract 
the NLO cross section for the production of Higgs plus jet(s) from the 
inclusive NNLO result.
The NNLO inclusive cross section in Eq.~(\ref{sigmaveto}) is evaluated by
using the results of 
Refs.~\cite{Catani:2001ic,Harlander:2001is,Harlander:2002wh}
\footnote{We include all the soft and virtual contributions and
the hard terms of the form $(1-z)^n$ up to $n=1$.
Higher powers of $(1-z)$ give very small effects \cite{Harlander:2002wh}.},
whereas the contribution $\Delta \sigma$
is evaluated by using the numerical program of Ref.~\cite{deFlorian:1999zd}.

In the following we present 
numerical results 
both at NLO and at NNLO.
These are obtained by using the
parton distributions of the MRST2001 set \cite{mrst2001},
with densities and 
QCD coupling 
evaluated at each corresponding order.
The MRST2001 set includes (approximate) NNLO parton densities. 
We fix $\mu_F=\mu_R=M_H$ (the scale dependence is studied in 
Ref.~\cite{Catani:2001cr}).
\begin{figure}[ht]
\begin{center}
\begin{tabular}{c}
\epsfxsize=11truecm
\epsffile{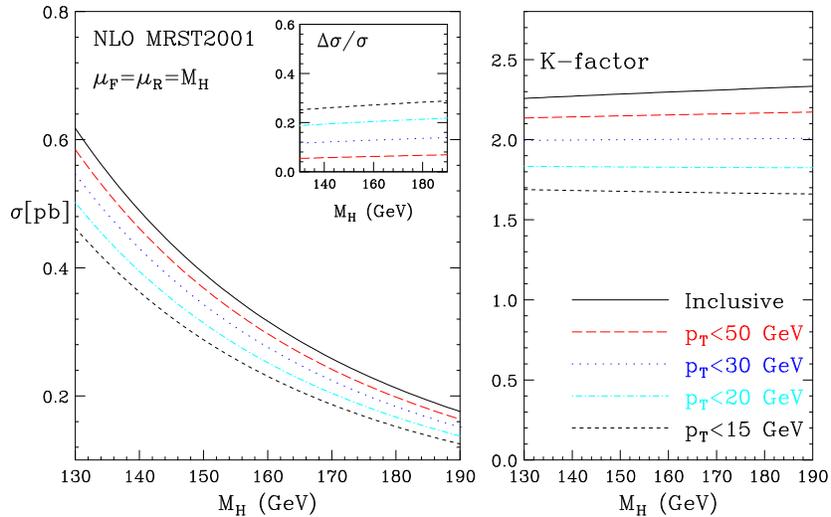}\\
\end{tabular}
\end{center}
\vspace*{-3mm}
\caption{\label{fig:tevnlo}{\em Vetoed cross section
and K-factors: NLO results at the Tevatron Run II.}}
\end{figure}
\begin{figure}[htb]
\begin{center}
\begin{tabular}{c}
\epsfxsize=11truecm
\epsffile{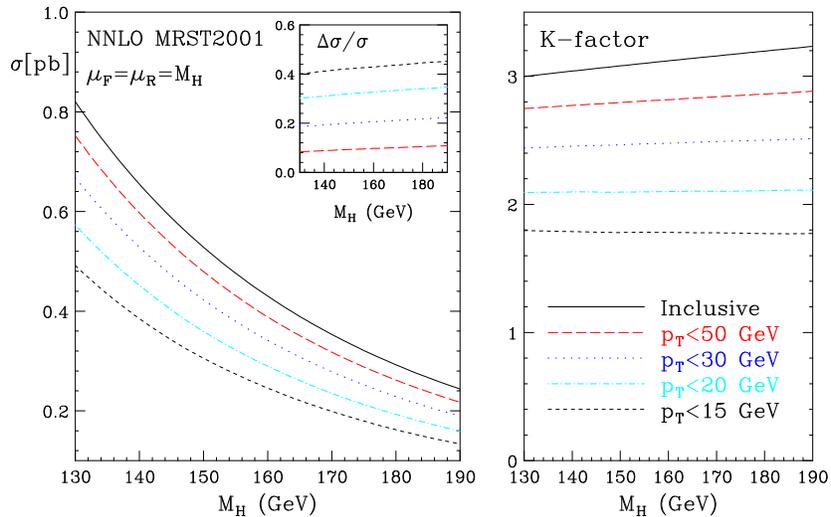}\\
\end{tabular}
\end{center}
\vspace*{-3mm}
\caption{\label{fig:tevnnlo}{\em Vetoed cross section and K-factors:
NNLO results at the Tevatron Run II.}}
\end{figure}

We first present results
at the Tevatron Run II.
In Fig.~\ref{fig:tevnlo} (Fig.~\ref{fig:tevnnlo}) we show the dependence of 
the NLO (NNLO)
calculation
on the Higgs mass for different values of 
$p_T^{\rm veto}$ (15, 20, 30 and 50~GeV).
The vetoed cross sections $\sigma^{\rm veto}(s,M_H^2;p_T^{\rm veto},R)$
and the inclusive cross section $\sigma(s,M_H^2)$ are given in the plots
on the left-hand side. The inset plots gives an idea of
the `loss' in cross section once the veto is applied,
by showing the ratio between the cross section difference $\Delta \sigma$ 
in Eq.~(\ref{sigmaveto}) and the inclusive cross section at the same
perturbative order.
As can be observed, for large values of the cut, say $p_T^{\rm veto}=50$~GeV,
less than $\sim 10\%$ of the inclusive cross section is vetoed. The veto effect
increases by decreasing $p_T^{\rm veto}$, but at NLO (NNLO) it is still 
smaller than $\sim 30\%$ ($\sim 40\%$)
when $p_T^{\rm veto}=15$~GeV.
On the right-hand side of Figs.~\ref{fig:tevnlo} and \ref{fig:tevnnlo}
we show the corresponding K-factors, i.e. the vetoed
cross sections normalized to the LO result,
which is independent of the value of the cut.
\begin{figure}[htb]
\begin{center}
\begin{tabular}{c}
\epsfxsize=11truecm
\epsffile{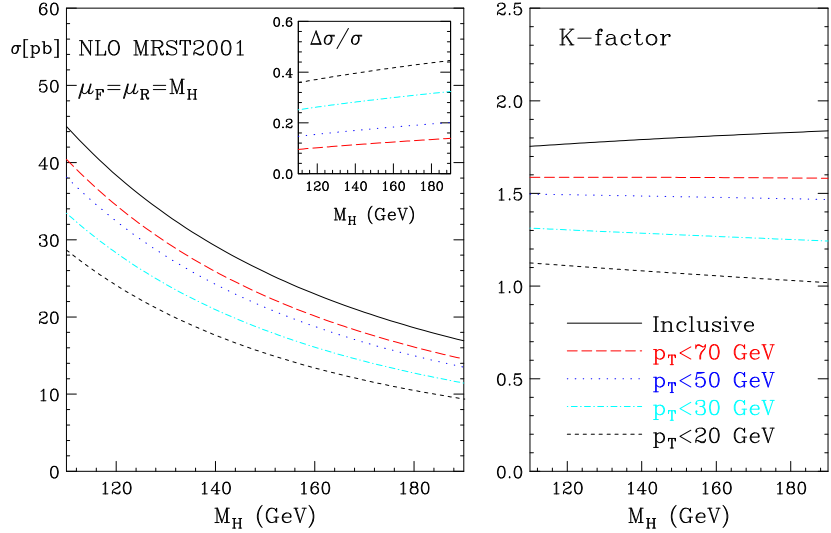}\\
\end{tabular}
\end{center}
\vspace*{-3mm}
\caption{\label{fig:lhcnlo}{\em Vetoed cross sections and
K-factors at NLO at the LHC. }}
\end{figure}
\begin{figure}[htb]
\begin{center}
\begin{tabular}{c}
\epsfxsize=11truecm
\epsffile{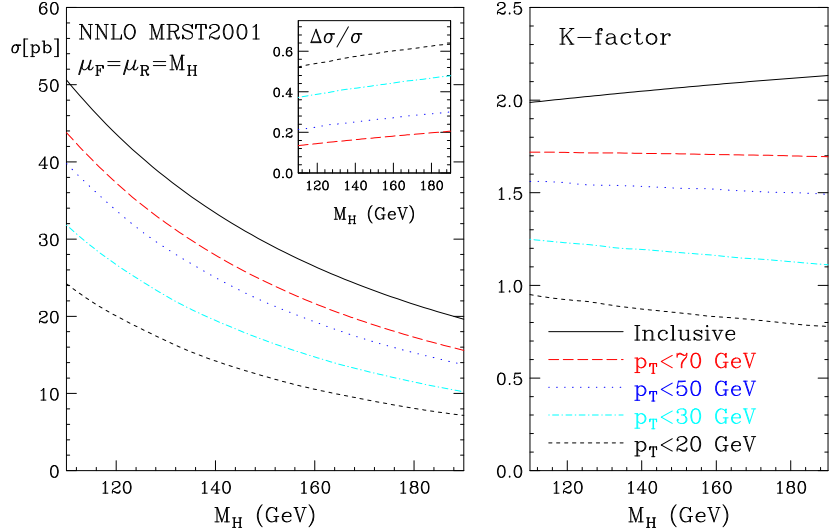}\\
\end{tabular}
\end{center}
\vspace*{-3mm}
\caption{\label{fig:lhcnnlo}{\em Vetoed cross sections and
K-factors at NNLO at the LHC.  }}
\end{figure}
 
The LHC results for $p_T^{\rm veto}=20$, 30, 50 and 70~GeV are presented in
Figs.~\ref{fig:lhcnlo} and \ref{fig:lhcnnlo}.
At fixed value of the cut, the impact of the jet veto, 
both in the `loss' of cross section and in the reduction of the K-factors,
is larger at the LHC than at the Tevatron.
For example, when $p_T^{\rm veto}=50$~GeV at the LHC we have
$\Delta \sigma/\sigma \sim 18\% (25\%)$ at NLO (NNLO).

The results presented above have a simple physical interpretation
\cite{Catani:2001cr}.
The dominant part of QCD corrections 
is due to soft and collinear radiation \cite{Catani:2001ic} (incidentally, this
justifies the use of the large-$M_{\rm top}$ approximation), and leads to
enhancement of the cross section.
The characteristic scale of the highest
transverse momentum $p_T^{\rm max}$ of the accompanying jets is 
$p_T^{\rm max}\sim \langle 1- z \rangle M_H$, 
where the average value 
$\langle 1- z \rangle = \langle 1 - M_H^2/{\hat s}\rangle$ of the distance
from the partonic threshold is small. As a consequence the jet veto procedure
is weakly effective unless the value of $p_T^{\rm veto}$ is
substantially smaller than $p_T^{\rm max}$.
Decreasing $p_T^{\rm veto}$,
the enhancement of the inclusive cross section due to soft radiation at higher
orders is reduced, and the jet veto procedure tends to
improve the convergence of the perturbative series.
At the LHC Higgs production is less
close to threshold than at the Tevatron and, therefore, the accompanying jets
are harder.
Thus,
at fixed $p_T^{\rm veto}$, the effect of the jet veto is
stronger at the LHC than at the Tevatron.

Note that the numerical results presented here (slightly) differ from those in
Ref.~\cite{Catani:2001cr}, because of the following three reasons.
$i)$ Here we include the exact $M_{\rm top}$-dependence of the Born cross
section $\sigma_0$, while in Ref.~\cite{Catani:2001cr} $\sigma_0$ was
approximated by its large-$M_{\rm top}$ limit. This affects the absolute
value of the cross sections, but not the ratios $\Delta \sigma/\sigma$ and the
K-factors. $ii)$ In Ref.~\cite{Catani:2001cr} we used the parton densities of
the MRST2000 set \cite{Martin:2000gq}.
The differences between the parton densities of Refs.~\cite{mrst2001}
and \cite{Martin:2000gq} lead to 
effects of $\sim 10\%$. 
$iii)$ The contribution of the NNLO hard terms \cite{Harlander:2002wh},
not included in Ref.~\cite{Catani:2001cr}, decreases the NNLO inclusive 
cross section by $\sim 7\%$ at the Tevatron
and $\sim 5\%$ at the LHC. Owing to the subtraction in Eq.~(\ref{sigmaveto}),
this NNLO decrease becomes relatively more important when
$p_T^{\rm veto}$ becomes small (i.e. when $\Delta \sigma$ increases).

\noindent {\bf Acknowledgments.}
This work was supported in part 
by the EU Fourth Framework Programme ``Training and Mobility of Researchers'', 
Network ``Quantum Chromodynamics and the Deep Structure of
Elementary Particles'', contract FMRX--CT98--0194 (DG 12 -- MIHT).

\end{document}